\documentclass[12pt,preprint]{aastex}

\newcommand\mdot{\rm \dot{M}}

\newcommand\msunyr{\rm M_{\odot}\,yr^{-1}}
\newcommand\be{\begin{equation}}
\newcommand\en{\end{equation}}

\newcommand\kms{\rm{\, km \, s^{-1}}}

\newcommand\etal{{\rm et al}.\ }

\shorttitle{IRS spectra of FU Ori Objects}
\shortauthors{Green et al.}

\begin{document}

\title{
Spitzer-IRS Observations of
FU Orionis Objects}

\author{J. D. Green\altaffilmark{1}, L. Hartmann\altaffilmark{2}, N. Calvet\altaffilmark{2},
 D. M. Watson\altaffilmark{1}, 
M. Ibrahimov\altaffilmark{3}, E. Furlan\altaffilmark{4}, B. Sargent\altaffilmark{1},
W. J. Forrest\altaffilmark{1}}

\affil{1. Department of Physics and Astronomy,
University of Rochester, Rochester, NY 14627;
2. Department of Astronomy, University of Michigan;
3. Ulugh Beg Astronomical Institute, Uzbek Academy of Sciences, Tashkent, Uzbekistan
4. Center for Radiophysics and Space Research, Cornell University, Ithaca, NY 14853-6801}

\begin{abstract}
We present 5-35 $\mu$m spectra, taken with the Infrared Spectrograph
(IRS) on the Spitzer Space Telescope, of five FU Orionis objects:
FU Ori, V1515 Cyg, V1057 Cyg, BBW 76, and V346 Nor.  All but V346 Nor reveal amorphous
silicate grains in emission at 10 $\mu$m and 20 $\mu$m, and
show water-vapor absorption bands at 5.8 and 6.8 $\mu$m and SiO or possibly methane absorption at 8 $\mu$m.
These absorption features closely
match these bands in model stellar photospheres --- signs of 
the gaseous photospheres of the inner regions of these objects' accretion disks.
The continuum emission at 5-8 $\mu$m is also consistent with such disks, and, for FU Orionis
and BBW 76, longer-wavelength emission may
be fit by a model which includes moderate disk flaring.  
V1057 Cyg and V1515 Cyg have much more emission at longer wavelengths than the others, perhaps 
evidence of substantial remnant of their natal, infalling envelopes. 
\end{abstract}
\keywords{}

\section{Introduction}

The FU Orionis variables are pre-main sequence systems which initially
were identified as having large ($\sim$ 5 mag), sudden outburts which typically decay in decades 
\citep[e.g.,][]{herbig77}.
FU Ori outbursts have been modeled 
as resulting from rapidly accreting disks in Keplerian rotation around pre-main
sequence stars, a model which explains their broad infrared Spectral Energy Distributions (SEDs) 
and the variation of spectral type and rotational velocity with the 
wavelength of observation \citep{hartmann96a}.  FU Ori disks
are thought to differ from those of most Classical T Tauri stars (CTTS)  
in that the heating of the disk is dominated by local accretion energy
release (viscous dissipation), at least in the inner disk regions, while
CTTS disks, although accreting, are thought to be heated mostly by radiation from the central
star \citep[e.g.,][]{kenyon87}.  Extended images show that FU Ori are typically
surrounded by remnants of the original 
molecular cloud material out of which the system formed, still
falling onto the disk \citep{herbig77,goodrich87}.

On the human time scale, FU Ori outbursts are rare events; 
there have only been $\sim$ 5-10 outbursts detected
in the past seventy years.  Compared to the local rate of star formation --
$\sim$ 0.02 $yr^{-1}$ within 1 kpc of the Sun \citep{miller79} -- they happen frequently enough 
that it is plausible to suggest that stars commonly experience such events.  Young
stars may experience them repeatedly, early in their development \citep{herbig77,hartmann96a}. 
If so, they may play
a fundamental role in early stellar protoplanetary-disk evolution:
the total mass accreted in these short-lived
events represents a significant fraction (10\% or more) of the final mass
of the developing young star.

Mid-infrared spectra from the Infrared Spectrograph \citep[IRS;][]{houck04}
on board the Spitzer Space Telescope
\citep{werner04} can provide crucial constraints on the 
structure of the inner few hundred AU of FU Ori systems.
These spectra may permit isolation of the contributions of
the cold envelopes and disks heated by viscous dissipation.
In this paper we report IRS observations of five
FU Ori objects -- FU Ori, V1515 Cyg, V1057 Cyg, BBW 76, and V346 Nor -- and discuss
their impact on models of FU Ori objects.

\section{Observations and data reduction}

\subsection{IRS observations}

Five FU Ori objects --
FU Ori, V1057 Cyg, V1515 Cyg, BBW 76, and V346 Nor --
were observed between December 2003 and May 2004 using Spitzer-IRS.  
Table \ref{tbl1} is a journal of the observations.
We operated the observatory in IRS Staring Mode, which
includes a high-accuracy pointing peak-up using the onboard Pointing Calibration and Reference Sensor
\citep{mainzer03}, and involves acquisition of spectra in pairs with the target displaced (nodded) along 
the spectrograph slit.  

All of these objects were observed in both orders of the IRS Short-Low spectrograph
(SL; 5.2-14 $\mu$m, $\lambda/{\Delta\lambda} \sim 90$).  BBW 76 was also observed in
both orders of the IRS Long-Low spectrograph (LL; 15-40 $\mu$m, $\lambda/{\Delta\lambda}
\sim 90$).
The other targets were observed in IRS Short-High (SH; 10-20 $\mu$m,
$\lambda/{\Delta\lambda} \sim 600$) and IRS Long-High (LH; 20-40 $\mu$m,
$\lambda/{\Delta\lambda} \sim 600$).  
We reduced the spectra using the IRS team's Spectral Modeling, Analysis, and Reduction Tool
\citep[SMART;][]{higdon04}.  We began with the ``droop'' data product from the Spitzer
Science Center IRS calibration pipeline version S11.0.2, consisting of 2-D spectral-spatial, 
non-flatfielded images.  The known wavelength calibration error in S11 was corrected before
extraction.  Before point-source spectral extraction, 
we corrected each pixel contained in the IRS team's mask of permanently bad 
or misbehaving pixels, by interpolation among its appropriately normalized neighbors. 
In IRS SL and LL observations both orders are always observed, but the point-source target 
is placed sequentially in each order, so the ``off'' orders provide sky-background measurements 
which we subtracted from the on-target spectra.  
We then extracted point-source spectra for each nod position using a column of variable
width scaled to that of the instrumental point-spread function.  
For our IRS SH and LH observations we simply extracted the
full slit, and did not subtract sky background emission, 
which is negligibly faint compared to the targets at these wavelengths.

In all cases, similarly-prepared spectra were produced from observations of 
two photometric standard stars, $\alpha$ Lac (A1 V) and $\xi$ Dra (K2 III).  We
then divided each target spectrum, nod position by nod position, by the spectrum of
the one of the standard stars and multiplied by the standard's template spectrum
\citep{cohen03}.  The final spectra are averages of the nod positions, except for V346 Nor,
for which only only one nod position was used.  In two cases, 
BBW 76 and V346 Nor,
slight pointing errors led to 20\% flux-density discrepancies between spectra from SL 
and the other (wider-slit) modules, which we resolved by increasing the SL signal 
by a scalar factor.  
As no significant narrow spectral lines were 
detected in the high-resolution data for V1057 Cyg, FU Ori,
and V346 Nor, we improved the signal-to-noise ratio in these spectra by convolution to 
lower spectral resolution, $\lambda/{\Delta\lambda} \sim 120$.  
We estimate the photometric accuracy of the final spectra to be approximately 5\%.

\subsection{Optical and near-IR photometry}

At any point in time, the continuum emission in our IRS spectra is a smooth extension
of the shorter-wavelength emission.  However,
because FU Ori objects display short-term variation in addition to the 
steady decay of the outburst, 
comparison to shorter-wavelength observations from the same epoch 
as the Spitzer-IRS observations is preferred.
Thus we employ UBVR observations of V1057 Cyg, FU Ori, and V1515 Cyg that were made at Maidanak Observatory 
throughout 2004, the continuation of a long monitoring campaign begun in 1981 for V1057 Cyg and V1515 Cyg,
and in 1984 for FU Ori.
Observations in this program before 2004 are available online 
\citep[see ][]{ibrahimov96,ibrahimov99,clarke05};
those for 2004, contemporaneous with our Spitzer-IRS observations, are presented 
in tables \ref{tbl2}, \ref{tbl3},
and \ref{tbl4}.
The data were taken with two 60-cm Zeiss reflectors and the 48-cm AZT-14 reflector equipped
with identical photon-counting photometers, reproducing the Morgan-Johnson system of photometry.
The observations were carried out using differences with nearby reference stars.  The RMS uncertainty of
a single measurement is $\Delta V = \Delta (V-R) = 0.015$ mag, $\Delta (B-V) = 0.02$ mag.
The RMS uncertainty in U-B was approximately 0.03 mag for FU Ori, and somewhat larger for V1057 Cyg
and V1515 Cyg.  The resulting average visible-UV
photometry for V1057 Cyg, FU Ori, V1515 Cyg and BBW 76 is displayed in table \ref{tbl5}.  

For BBW 76, we used the visible-UV data compiled by \citet{reipurth02} to
determine average decay rates between 1983 and 1990 for UBV photometry and 
linearly extrapolate values for 2004.  The decay rate in
the B and V-band was $\sim$ 0.02 mag $yr^{-1}$, while the U-band decay rate was
$\sim$ 0.003 mag $yr^{-1}$.  Next we calculated the decay at near-infrared wavelengths
($J$, $H$, and $K_S$ bands) by comparison of 1983 and 1998 (2MASS) observations.  
This yielded a decay rate of $\sim$ 0.026 mag $yr^{-1}$.
We assumed this rate to apply to $R$ as well, and extrapolated to 2004 from the 
1994 data in \citet{reipurth02}. 

Contemporaneous near-infrared photometry is not available for any of our targets, but comparison of older 
observations with those in the 2MASS point-source catalogue (PSC), as in BBW 76 above, indicate that 
the extrapolation from the 2MASS epoch ($\sim$ 1998) to the IRS epoch is negligibly small. Thus for 
V1057 Cyg, V1515 Cyg, and BBW 76 we adopt the 2MASS PSC $J$, $H$, and $K_S$ magnitudes.  
FU Ori is too bright to have been included in the 2MASS PSC; instead we used the 1989 near-infrared 
observations by \citet{kenyon91}.  Though these observations are significantly older than 2MASS, 
the M-band ($\lambda = 4.8~\micron$) flux density matches closely at the shortest wavelengths in the IRS
spectrum, indicating that FU Ori has not varied significantly at these wavelengths over
the past fifteen years.

\section{Observational results}

The IRS spectra of our five program objects are displayed in Figures \ref{fig1},
\ref{fig2}, and \ref{fig3}.
Four of the five objects ---
FU Ori, BBW 76, V1515 Cyg, and V1057 Cyg --- have broad silicate emission features 
(peaking at $\lambda$ = 10 and 18 $\micron$), and steeply-rising
continua shortward of the silicate features. At the shorter wavelengths 
the spectra of these four objects are similar. 
In each case, 
the 5-9 $\micron$ continuum (Figure \ref{fig2}) is 
modulated by broad absorptions at $\lambda$ =  5.8, 6.8 and 8.0 $\micron$. 
Based on comparisons with
IRS spectra of low-mass stars \citep{roellig04}, and model photospheres of 
low-mass stars and brown dwarfs \citep[e.g.,][]{allard00}, we identify the two shorter wavelength 
features with a collection of rotation-vibration bands in gaseous $\rm{H}_2\rm{O}$. \citet{calvet91}
previously noted that features between 1.4 and 2.4 $\micron$ match models of water vapor absorption.
Tentatively we 
suggest that the feature at 8 $\micron$ is the fundamental rotation-vibration band of gaseous SiO,
but note that it could potentially be due to or influenced by absorption by methane (${\rm CH}_4$) as well. 

Suppose that FU Ori disks exhibit gaseous features
at effective temperatures $> 1400$~K, but only featureless dust continuum
at lower effective temperatures.  The continuum corresponds to the hottest dust temperatures
seen in CTTS disks \citep{muzerolle03}.  In this case the dust opacity dominates the
gaseous opacity at temperatures $\lesssim 1400$ K.  From the standard
steady, blackbody FU Ori disk model with a large outer radius \citep{kenyon88},
we estimate that almost half of the flux at 6 $\mu$m, and about a third of the flux
at 8 $\mu$m, is contributed by annuli hotter than 1400 K.  
This model predicts that we 
should detect gaseous features at these wavelengths in FU Ori objects,
diluted by dust continuum emission from cooler disk regions.
The appearance of strong emission due to dust at just 
slightly longer wavelengths (Figure \ref{fig1}) warns us that 
the 5-9 $\micron$ continuum 
emitted by dust grains in cooler parts of the accretion disk is probably still significant here, 
as previously found in the spectra of CTTSs \citep{furlan05}.
However, neither suitable models nor observed spectra 
of disk photospheres at mid-infrared wavelengths yet exist for assessment
of the dust-continuum dilution of the gaseous features. 
The spectra of FU Ori, BBW 76, V1515 Cyg, and V1057 Cyg are not as similar at 
the longer IRS wavelengths: the Cygnus objects are quite a bit redder at 
$\lambda~=~15-40 \micron$ than the other two, 
indicating a greater predominance in these objects of emission by colder dust.

\subsection{V346 Normae}

In contrast to the others, V346 Nor shows strong silicate absorption, consistent
with its large visual extinction; \citet{graham85} estimate $A_V \sim 6.2$ mag.
Figure \ref{fig3} is a comparison of IRS spectra for V346 Nor and a typical Class I 
young stellar object, IRAS 04016+2610 \citep{watson04}.  The spectra are 
quite similar, and thus the identity of the features is the same as 
in Class I objects, namely  water ice ($\lambda~=~6.0~\micron$), carbon dioxide ice
($\lambda~=~15.2~\micron$), a feature usually associated with methanol ice
($\lambda~=~6.8~\micron$), and amorphous silicates ($\lambda~=~10, 18~\micron$).

\subsection{V1057 Cygni}

The 10 $\mu$m silicate
feature of V1057 Cyg appears different from the others.  As we show later, 
after correction by the adopted $A_V = $ 3.7,
using a reddening curve from \citet{savage79}
the difference vanishes (Figure \ref{fig4}). Thus the appearance of the 
feature is due to selective extinction, rather than difference in dust composition. 
After subtracting a continuum derived from the 6-8 $\mu$m region, and deriving an emissivity
based on a single temperature model for the dust grains \citep{sargent06}, we find that the 
peak of the silicate
emission is at 9.5 $\mu$m, which is indicative of amorphous silicates of pyroxene composition.
This is consistent with dust composition common in CTTS disks \citep{demyk00,sargent06}, 
and within the usual range of composition of interstellar dust.
We have not attempted such a fit to the silicate features
of the other objects in the study; their silicate features have approximately 
the same peak wavelength and width as those of V1057 Cyg (see Figure \ref{fig4}).
They all appear smooth and should yield composition similar to V1057 Cyg. 

\subsection{V1515 Cygni}

\begin{figure}[ht]
\plotone{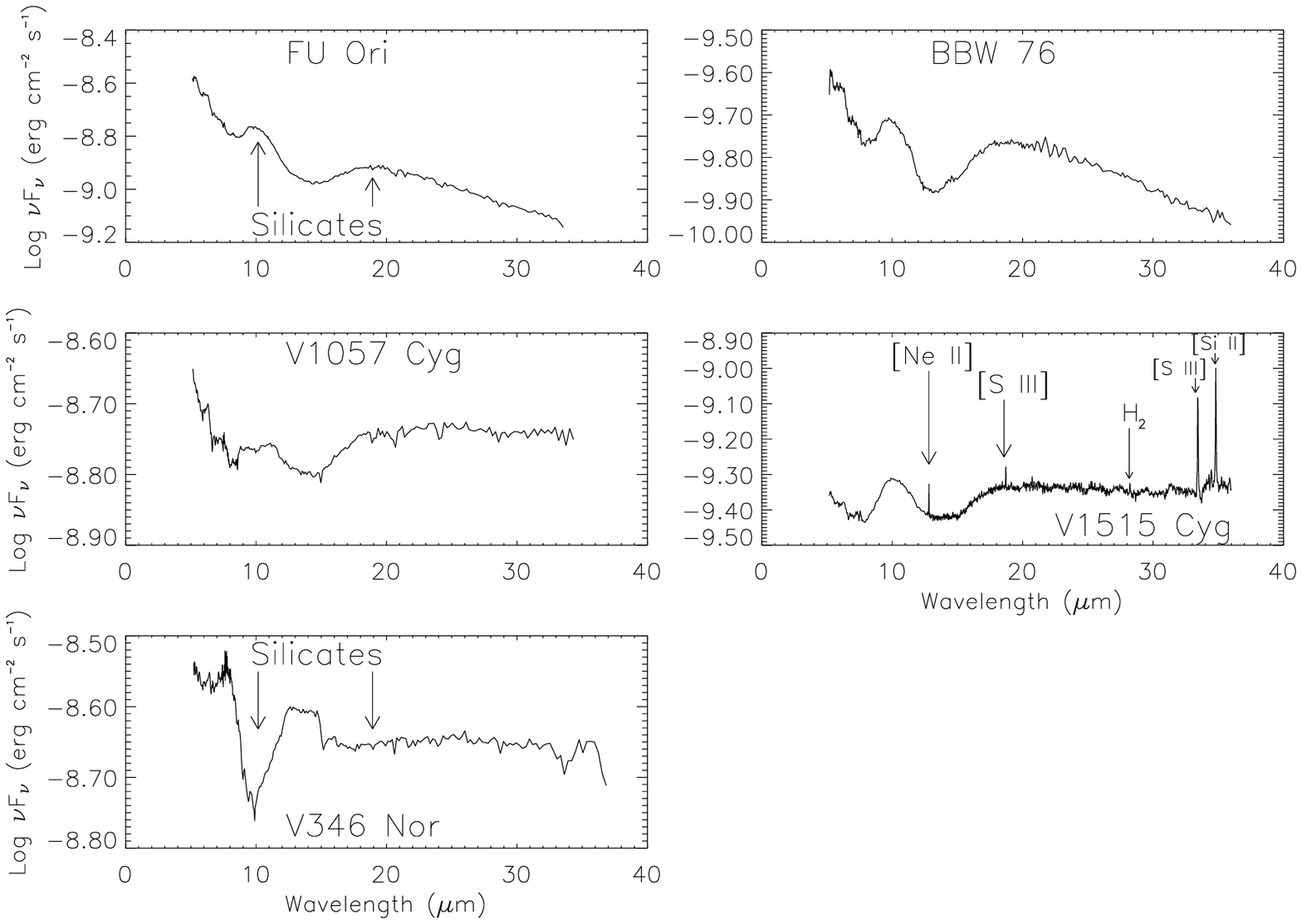}
\caption{Observed IRS spectra of FU Orionis variables, uncorrected for reddening.
The data for BBW 76 and V1515 Cygni
are shown at full resolution: BBW 76 was a low resolution observation, and we show the
high resolution data for V1515 Cygni in order to display the spectral lines.  The data
for the other three objects have been rebinned to low resolution.  As discussed in the
text, the ionic emission lines in V1515 Cyg arise in the foreground or the background.}
\label{fig1}
\end{figure}

\begin{figure}[ht]
\plotone{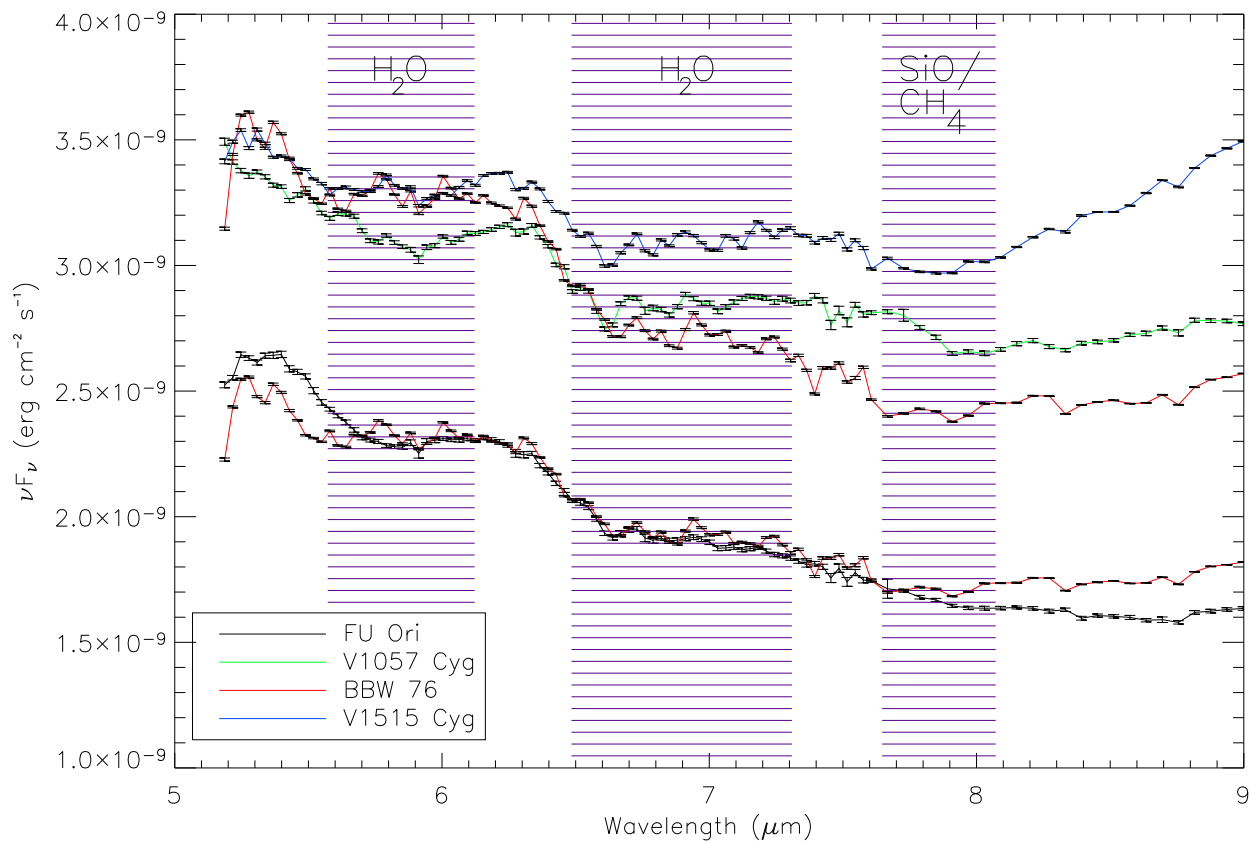}
\caption{Comparison of the 5-8 $\mu$m region of the four FU Ori objects showing gaseous
absorption features: $H_2O$ absorption bands 5.8 and 6.8 $\mu$m and an SiO band 
at 8 $\mu$m.  The spectra have been scaled, and BBW 76 is shown twice at
different scales for comparison to the sources with more emission at longer wavelengths
(V1057 Cyg, V1515 Cyg), and to the source it most closely resembles (FU Ori).  The error bars
in the plot are derived from the difference of the two nod positions.}
\label{fig2}
\end{figure}

\begin{figure}[ht]
\plotone{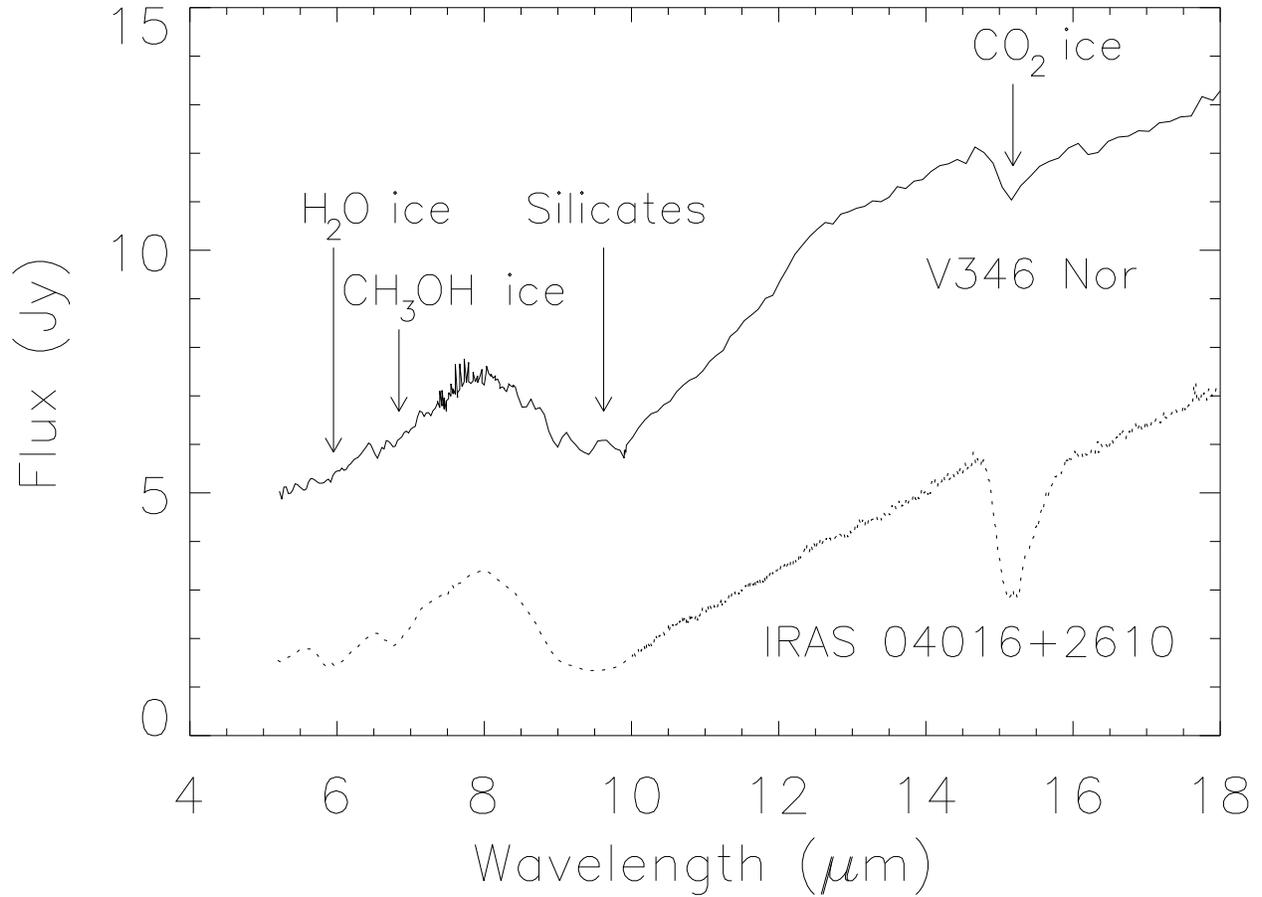}
\caption{Comparison of the 5-18 $\mu$m region of V346 Nor to the Class I YSO IRAS 04016+2610
\citep{watson04}.
The spectra are unscaled, although V346 Nor has been rebinned to low resolution.  The spectrum of
V346 Nor is considerably more noisy despite being brighter, because of significant telescope pointing
errors.}
\label{fig3}
\end{figure}

As is common in the Cygnus clouds, 
the line of sight toward V1515 Cyg intersects a great deal of foreground and background nebulosity,
and as a result the spectrum exhibits many spectral features from ions, molecules and small dust grains. 
Most prominent are the infrared bands identified with polycyclic aromatic hydrocarbons (PAHs),
at $\lambda$ =  6.2, 7.7, 8.6 and 11.3 $\micron$. This emission is removed almost 
completely by the off-order sky subtraction carried out for SL spectra during data reduction, and 
thus does not appear in the spectrum shown in Figure \ref{fig1}. That it subtracts away so precisely is 
an indication that the emitting material is neither associated with, nor excited by, V1515 Cyg. 
The same is true for [Ne II] emission at $\lambda$ =  12.8 $\micron$. For wavelengths longward of 14
$\micron$, observed only with SH and LH, we lack a nearby blank-sky spectrum for subtraction, and thus 
spectral lines from extended nebulosity appear in Figure \ref{fig1}: [S III] (18.7 $\mu$m
and 33.4 $\mu$m), $H_2$ (28.2 $\mu$m), and [Si II] (34.8 $\mu$m).  
We expect that all these lines therein would subtract away as precisely as [Ne II], the SH profile of
which we leave in Figure \ref{fig1} for comparison.

\section{Discussion}

\subsection{Emission by the hot inner disk: comparison to simple models} 

\begin{figure}[ht]
\plotone{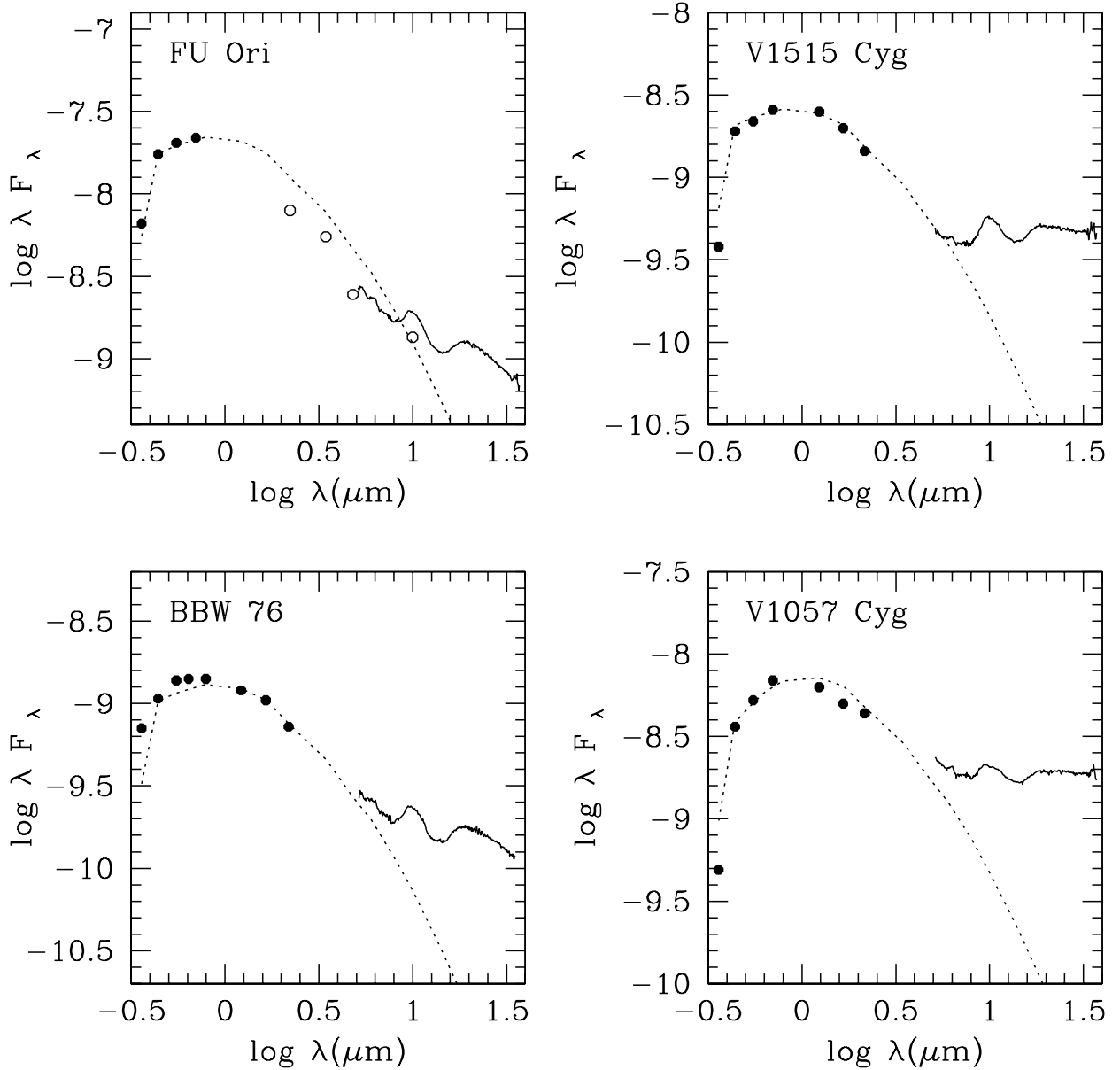}
\caption{Dereddened IRS spectra (solid line) and older optical and near-IR ground-based data (circles); the dashed line
is a fit from the steady accretion model 
using standard photometric colors down to 3000 K from
Kenyon \& Hartmann 1995 (Table A5) with fit parameters listed in Table 6.  Below 3000 K, blackbodies are
used.  The value of $A_V$ for
each object is listed in Table 5.  The filled optical data circles are observations made at
Maidanak Observatory during 2004; the filled near-IR data circles are 2MASS observations circa 1998; the open circles
are 1989 KLMN bolometer data from KH91.}
\label{fig4}
\end{figure}

To compare the SEDs of the four moderately-reddened
FU Ori objects with disk models, it is necessary to make
extinction corrections, as most of the disk luminosity is
radiated at visible wavelengths.  Extinction corrections are uncertain because these
systems do not have single, well-defined effective temperatures;
this leads to larger uncertainties in SED shapes and system luminosities.
We apply extinction corrections to the ground-based photometry
\citep{savage79} and the IRS spectra, assuming that the optical
depth at 9.7 $\mu$m $\tau_{9.7}$ = $A_V$/18 \citep{schutte98,draine03}.
These extinction values are reasonably consistent
with previous estimates in the literature but may differ by as much as
0.2 to 0.5 magnitudes in $A_V$.  This in turn leads to an uncertainty in
the peak of the SED, generally around 1 $\mu$m or a little less, of
order 20-30\%, with a similar uncertainty in the system luminosity
for a given distance.  The resulting, corrected spectra appear in Figure \ref{fig4}.

In Figure \ref{fig4},
we also show SEDs for steady accretion disk models assuming no inclination along the line-of-sight.
The most common steady accretion disk models \citep[e.g.,][]{hartmann98} assume the disk radiates as a blackbody at a 
temperature that varies with radius R according to

\begin{equation}
T_d^4 ~=~  
\frac{3GM\dot M}{8\pi R^3 \sigma} \left[ 1- \left( \frac{R_i}{R} \right)^{1/2} \right]
\end{equation}
The maximum temperature of the disk is then given by 

\begin{equation}
T_{max} ~=~
0.488 \left(\frac{3GM\dot M}{8\pi R_i^3 \sigma} \right)^{1/4}
\end{equation}
which occurs at \begin{equation} R_{max}= \frac{49}{36}R_i \end{equation}
Thus the total luminosity can be calculated:

\begin{equation}
L_\nu ~=~ 
\int_{R_{in}}^{R_{out}}{\pi S_\nu T_d(R)2\pi R dR}
\end{equation}
assuming no inclination
along the line-of-sight .
Rather than using blackbodies ($S_\nu~=~B_\nu(T)$)
we used photospheric emission computed using 
the photometric colors of standard stars taken from Table A5
of \citet{kenyon95} to provide a slightly more realistic
estimate of real disk flux density at short wavelengths (blackbodies 
are used at long wavelengths and for all wavelengths at disk 
temperatures below 3000 K).  
The same disk 
model is used for FU Ori, BBW 76, and V1515 Cyg, slightly scaled 
in luminosity, for ease of comparison; for V1057 Cyg we have used 
a disk model with a slightly lower $T_{max}$ (table \ref{tbl6}). 
Steady disk models are in fair to good
agreement with the observations shortward of about 6 $\mu$m, but at longer
wavelengths there is more flux at longer wavelengths than would be predicted
by of the models.  

Disks heated from the outside will have higher temperatures in
surface layers than deeper in, a property which produces 
silicate emission features \citep{calvet92}.  In contrast,
FU Ori objects have optical and near-infrared absorption features which are
interpreted as resulting from internal heating by viscous dissipation
\citep{hartmann85,hartmann87a,hartmann87b}.
All the FU Ori objects in Figure \ref{fig4} show silicate emission
features at 10 and 18 $\mu$m, whereas at shorter wavelengths absorption
bands are observed.  It therefore seems likely
that the spectrum longward of about 8 $\mu$m or so should be considered
in terms of irradiation, either of a flared disk or an envelope.
We therefore first analyze the short-wavelength IRS SEDs in terms of accretion
disk models and then consider irradiation of disks and
envelopes for explaining the longer wavelength region
in section \ref{4.3} below.

\subsection{Disk models with radially-variable accretion rate}

\begin{figure}[ht]
\plotone{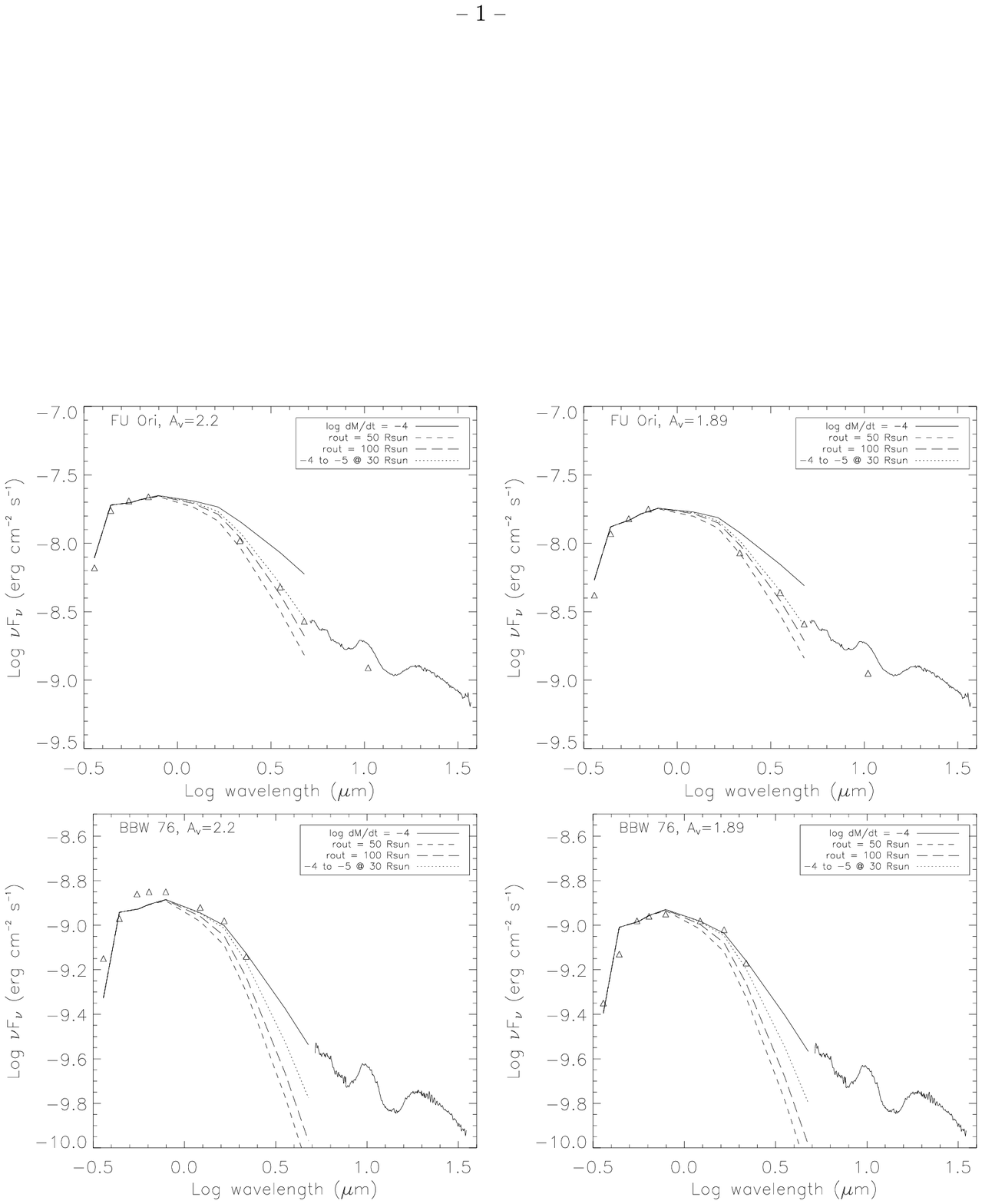}
\caption{Comparison of different disk models of FU Orionis with varying conditions in the outer disk.
The solid line represents a disk with a constant accretion rate of $10^{-4}$ $M_{\sun}$ $yr^{-1}$.  The dotted line
represents a model where the accretion rate is decreased to $10^{-5}$ $M_{\sun}$ $yr^{-1}$ beyond 30 $R_{\odot}$.
The longer dashed line represents a disk which is truncated beyond 100 $R_{\odot}$, and the disk represented by
the shorter dashed line is truncated beyond 50 $R_{\odot}$.
Upper left: The model shown for FU Ori uses the values given in table \ref{tbl6}.
Upper right: The same comparison after applying a lower extinction correction, assuming $A_V = 1.89$, and
reducing $M \dot{M}$ from 1.5 to 1.1.  For this model, $T_{max}$ = 7140 K.
Lower left:  The same comparison for BBW 76.  The parameters are taken from table \ref{tbl6}.
Lower right: $M \dot{M}$ has been reduced from 0.72 to 0.65, consequently reducing $T_{max}$ to 7140 K.
The optical photometry in the case of BBW 76 does not
match the disk model as well as it did for FU Ori, but the extended steady disk model which passes through
the 2MASS JHK data connects well to the IRS spectrum.}
\label{fig5}
\end{figure}

FU Ori stands out in Figure \ref{fig4} because its SED appears to be a bit narrower,
with less infrared emission, than a standard steady disk model.  This had
been found before (e.g., KH91, Figure 5) without
explicit comment.  Of course, as FU Ori objects vary, they cannot be represented
precisely by steady accretion disks.  Moreover, thermal instability models
of FU Ori outbursts, which can explain rapid rise times \citep{bell94,bell95},
predict that only an inner region of the disk participates
in the outburst; this implies that, beyond a certain radius, the accretion
rate is lower, reducing the amount of long-wavelength emission from outer,
cooler regions.  Thus we have also constructed simple disk models of this type, with
results as shown in Figure \ref{fig5}.

The upper left panel of Figure \ref{fig5} shows the FU Ori SED with various disk models.
Since the \citet{bell94}
and the \citet{bell95} FU Ori models predict a much higher accretion rate
in outburst within $\sim$ 20 $R_{\odot}$ than outside this radius (where the 
accretion rates tends toward the supplied infall rate), a first approximation
might be to assume that the outer disk accretion rate is so much lower that
the disk is effectively truncated at some radius.  However, even within the
limitations of the data, it is clear that disk models with substantial
accretion rates must extend well past this radius to explain the observations.
Steady  disk models need to extend out to $\sim$ 100 $R_{\odot}$ or more
to explain our FU Ori spectra.  A more realistic approximation in the thermal instability models would be to
assume a lower but non-zero accretion rate, steady disk temperature distribution
outside of the radius at which the thermal instability is triggered.  As shown
in the upper left panel of Figure \ref{fig5}, a model in which the accretion rate is arbitrarily
dropped by a factor of ten at $\sim$ 30 $R_{\odot}$ does a reasonably good job
of joining to the IRS spectrum.  Again, this requires an accretion rate
in the disk which is two orders of magnitude higher than typical T Tauri
rates \citep[$\sim$ $10^{-8}$ $\msunyr$ in Taurus;][]{calvet04}, out to 100
$R_{\odot}$ or more.

These models of FU Ori have higher maximum temperatures,
7710 K, than the 7200 K adopted by \citet{kenyon88}.  Part of the difference may
be due to the adoption by \citet{kenyon88} of a constant inner temperature 
interior to $1.36 R_*$, whereas our steady accretion disk models have the
peak temperature at
this radius and then decline inwards assuming a slowly-rotating central star;
\citep[see ][]{hartmann98}.  We have also examined
the effect of dropping the maximum temperature to correspond to a decrease
in extinction of $\Delta A_V = 0.3$ in visual magnitudes (certainly within the uncertainties),
also shown in Figure \ref{fig5}.
Although the new model disks are less luminous,
with a lower $T_{max}$ of 7140 K, there is little difference in the model comparison
at long wavelengths, because the SED shifts slightly
toward longer wavelengths, compensating for the decrease in system luminosity.
Thus the conclusion that the disk of FU Ori must be rapidly accreting out
to of order 100 $R_{\odot}$ or more
appears to be insensitive to plausible changes in the extinction estimate.
 
Also shown in Figure \ref{fig5} is a comparison of the SED of BBW 76 with
a steady disk model similar to that used for FU Ori - that is, a disk model
with the same maximum temperature, but with a slightly smaller inner radius
(Table \ref{tbl6}).  The optical photometry does not
match the disk model as well as in FU Ori. 
On the other hand, the extended steady disk model which passes through
the recent 2MASS data points joins fairly well to the IRS spectrum.
Again, the conclusion is that BBW 76 has relatively high accretion rates out to
radii of order 100 $R_{\odot}$ or more.

As shown in Figure \ref{fig4}, the steady disk model which passes through the 
2MASS photometric points in V1515 Cyg extrapolates very well to the 
short-wavelength end of the IRS spectrum.  This may be somewhat misleading,
as there is evidence for a contribution from the dust disk or envelope
in the 6-8 $\mu$m region in addition to the contribution from the steady
accretion disk (see following subsection).  The IRS spectrum of V1057 Cyg
clearly falls well above the extrapolation of a steady disk model, but this
excess is probably due to the envelope or disk regions responsible for the
far-infrared excess and silicate emission (see next subsection).

\subsection{Long-wavelength dust emission and envelopes}\label{4.3}

\citet{adams87} suggested that FU Ori might have a
tenuous spheroidal envelope, based in part on IRAS observations.
KH91 combined IRAS data with other IR observations
to study the longer-wavelength regions in FU Ori objects, with
special emphasis on FU Ori, V1057 Cyg, and V1515 Cyg, and found that
the infrared spectrum of FU Ori might be fit by a disk without an envelope.
KH91 concluded that V1057 Cyg and V1515 Cyg needed to have an additional
dusty envelope subtending a significant solid angle, which contributed strongly at wavelengths
$\gtrsim 10$ $\mu$m.  KH91 suggested that this dusty structure represented infalling
material which replaced the disk mass accreted in each outburst as a way
of obtaining multiple outbursts in the lifetime of each FU Ori object.
Further investigations of model envelopes were carried out by \citet{turner97}.
The IRS data provide an opportunity to reexamine more precisely the question
of disk vs. envelope irradiation.

\begin{figure}
\plotone{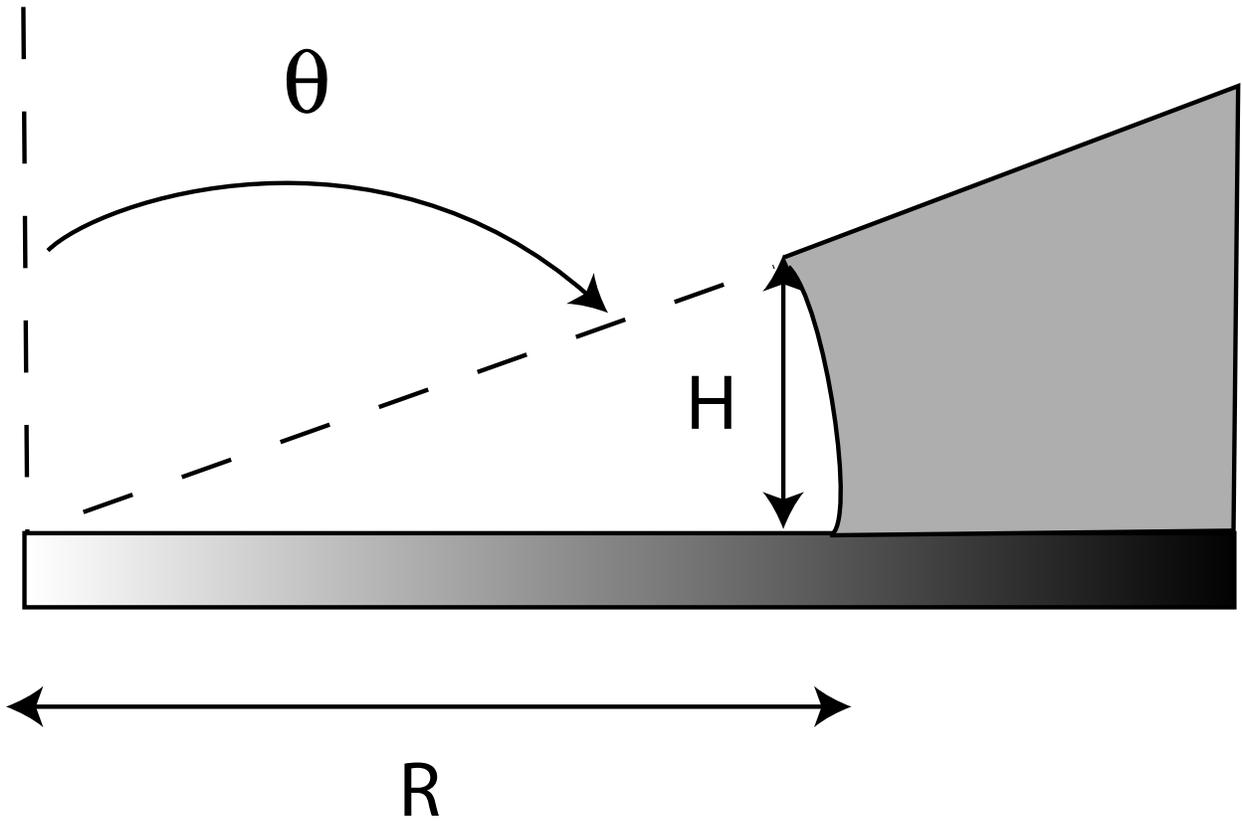}
\caption{Geometry of the simple disk plus envelope model (see text).}
\label{fig6}
\end{figure}

It may not be obvious that an envelope need be invoked, as T Tauri
disks can have SEDs that flatten out beyond $\lambda \gtrsim 10$ $\mu$m
\citep[e.g.,][]{kenyon87,chiang97,dalessio99}.
However, in the case of T Tauri disks, the irradiating source is the
star, not the inner disk, and this makes a crucial difference.  
Consider the geometry shown in Figure \ref{fig6}, where for simplicity we have
indicated a wedge of material which captures all the radiation emitted
at angles $\theta > \theta_m$.  In a more realistic disk/envelope combination
or a flared disk, this absorbed radiation would be distributed over a wider range of
radii, such that the height of the disk or the envelope 
at $R$ is $H$. Assuming an isotropically-emitting central source, the
fraction of the central luminosity $f_*$ absorbed within $R, H$ is simply
the fractional solid angle subtended by the envelope or disk region
as seen from the central source \citep[e.g.,][]{hartmann98},
\begin{equation}
f_* = \int_{\theta_m}^{\pi/2} \, d\theta \sin \theta  = \cos \theta_m\,, 
\label{eq:fstar}
\end{equation}
where $\theta_m = \arctan (R/H)$.
On the other hand, if the radiating source is a flat disk, even if the
emergent intensity is assumed to be isotropic, the emitted {\em flux}
is peaked in the direction perpendicular to the disk, due to the projection
of the emitting area in the line of sight.  For a geometrically flat disk
with luminosity $L_d$, the {\em apparent} luminosity at a viewing angle
$\theta$ is 
\begin{equation}
L(app) = 2 L_d \cos \theta \label{eq:ldisk}
\end{equation}  
Thus, the fraction of the radiation from a flat inner disk, much smaller
in radius than $R$, that can be
intercepted by the structure in Figure \ref{fig6} is
\begin{equation}
f_d = \int_{\theta_m}^{\pi/2} \, d\theta \, 2 \cos \theta \sin \theta  
= \cos^2 \theta_m\,. \label{eq:fdisk}
\end{equation}

A typical, strongly-flared T Tauri disk might have an aspect ratio
of about $H/R = 0.2$ and thus could intercept 
about 20\% of the light from its central star (equation \ref{eq:fstar}).
On the other hand, the same disk around the flat inner irradiating
disk can intercept only about 4\% of the inner disk luminosity
(equation \ref{eq:fdisk}).  Thus, large long-wavelength excesses 
due to irradiation by a flat disk require dusty structures which
subtend much larger solid angles than would be the case if the central
object radiated isotropically.

Thus, unlike a star, the apparent luminosity of a flat disk
can be as much as a factor of two larger than its true luminosity
if seen pole-on (equation \ref{eq:ldisk}).  In particular, V1515 Cyg
and V1057 Cyg have much smaller values of projected rotational velocity,
$v \sin i$, than those of FU Ori or BBW 76 \citep{hartmann85,hartmann87a,hartmann87b,reipurth02},
which suggests that the former two objects are observed
much closer to pole-on.  This suggestion is reinforced by the analysis
of \citet{goodrich87}, who argued for similar low viewing inclinations
for V1057 Cyg and V1515 Cyg on the basis of outflow envelope morphology.
To the extent that the long-wavelength excess is produced in an envelope
which emits more isotropically than the inner disk, the envelope will appear
to absorb a smaller fraction of the total luminosity in a pole-on system
than in reality.

We now consider the case of V1057 Cyg.  The disk model shown in Figure \ref{fig4}
corresponds to a total flux received at the Earth of 
$\sim$ 1.5 $\times 10^{-8}$ ${\rm erg \, cm^{-2}\, s^{-1}}$.
The total observed flux in the IRS range is about 3.6 $\times 10^{-9}$ 
${\rm erg \, cm^{-2}\, s^{-1}}$.  If we assume that all of the flux
at 5 $\mu$m is due to the accretion disk and not to the dusty structure
responsible for the long-wavelength excess, and extrapolate this
accretion disk flux as $\nu F_{\nu} \propto \nu^{4/3}$ to
longer wavelengths or lower frequencies, then the
excess above this disk emission is 
about 2 $\times 10^{-9}$ ${\rm erg \, cm^{-2}\, s^{-1}}$.
However, as discussed above, there is evidence for dust excesses
contributing at short wavelengths in V1057 Cyg; in addition,
the IRS SED shows no signs of turning down at the longest
wavelengths, so we have underestimated the total long-wavelength emission.
We adopt $\sim$ 3 $\times 10^{-9} {\rm erg \, cm^{-2}\, s^{-1}}$ as an
estimate of the total long-wavelength excess; this is 20\%  
of the apparent disk luminosity.  Using equation (\ref{eq:fdisk}),

$\cos (\theta_m) \sim (0.2)^{1/2}$ or $\theta_m \sim 63^{\circ}$,
so $H/R \sim \tan (90 - \theta_m) \sim 0.5$.  This is considerably larger
aspect ratio than for most known T Tauri disks.
In addition, the transition in the height of the absorbing dust structure
must be fairly abrupt to explain the SED.  We conclude, as did KH91,
that there is an additional dusty envelope to explain the far-infrared excess.
The true aspect ratio could be even larger if
we are viewing V1057 Cyg nearly pole-on and the envelope radiates more
isotropically than the inner disk.

At the other extreme, consider FU Ori.  Assuming that the accretion
disk emission dominates at 5 $\mu$m, which seems reasonable, and extrapolating
this steady disk spectrum as before, the long-wavelength excess is approximately 
2\% of the total system luminosity.  Even if we assume that all of the
flux in the IRS range arises from an excess component, this constitutes no
more than 4\% of the total luminosity.  According to equation (\ref{eq:fdisk}),
this excess in principle can be explained by absorption in a modestly flared
disk with $H/R \lesssim 0.2$.  We therefore reinforce the conclusion of KH91
that the SED of FU Ori can be explained by a flared disk without significant
envelope contributions in the IRS wavelength range.

V1515 Cyg has a SED which is quite similar in shape to that of V1057 Cyg 
(Figure \ref{fig4}), but with
a bit smaller long-wavelength excess.  The excess over the extrapolation of the 
steady disk model is about 10\%, and the total flux in the IRS range is about 14\%
of the total luminosity; again, the SED shows no sign of turning down at long
wavelengths so there must be significant longer-wavelength flux that we are
not including.  Adopting an estimated excess of about 12\% implies
$H/R \sim 0.37$, larger than typical of T Tauri disks.  In addition, V1515 Cyg
has an extremely low $v \sin i$, implying a very low angle of viewing.
As discussed above, this suggests that we are underestimating the relative 
magnitude of the long-wavelength excess by overestimating the central luminosity,
by perhaps as much as a factor of two; this would increase the required
$H/R$ even more, similar to that of V1057 Cyg.  We therefore conclude that
V1515 Cyg must also have a dusty envelope which dominates the long-wavelength
infrared emission.

BBW 76 represents an ambiguous case.  The shape of its SED is more similar to
that of FU Ori than to V1057 Cyg (Figure \ref{fig4}), but with a considerably larger
excess.  The excess above the steady disk model is about 5\% of the total
luminosity; this is very sensitive to the disk model assumed, as the
the total emission in the IRS range is about 10\% of the total.  
If we conservatively estimate a total excess of 6\%, then $H/R \sim 0.25$.
Thus it is not clear whether some additional, perhaps relatively optically-thin
envelope needs to be added to a flared disk to explain the IRS SED.

\citet{bell94}, \citet{bell95}, and \citet{bell97} constructed detailed
models for FU Ori objects based on specific mechanisms for the outburst
and parameters.  They pointed out that the inner disk might exhibit
departures from purely flat geometry.  However, it seems unlikely that
this will have a major effect on the above analysis; any regions which
have ``walls'' which are more face-on to the envelope or disk are
not likely to be large in extent, and we have neglected limb-darkening,
which will reduce the flux at large $\theta$ even more than
we have assumed.  Bell \etal also
constructed a number of disk and envelope models to explore the long-wavelength
SEDs of FU Ori objects; however, these models have many parameters and
it is not entirely clear how to apply their results to our new observations.   
The above analysis has the virtue of being minimally model-dependent while
illustrating the basic geometry.

Our suggestion of envelopes that intercept significant amounts of radiation
from the central disk in V1057 Cyg and V1515 Cyg is supported by recent
observations in the K-band with the Keck interferometer \citep{millan06},
who found that these objects are significantly more extended than
predicted by the accretion disk model.  Millan-Gabet et al. estimated that
their observations could be explained if 3 - 12\% of the K-band flux in V1057 Cyg and 
5 - 14\% of the K flux in V1515 Cyg arises from an extended source that
is well-resolved (i.e., bigger than 1 AU or so), as the KH91 envelope
model would predict.  The approximate amount of K-band excess, on the order of 5-10\%
of the central source K luminosity, is in reasonable agreement with the intercepted
fractions of disk luminosity on the order of 10\% derived above, assuming a 
moderately large ($\sim 0.3 - 0.5$) effective albedo at K.   
In contrast, FU Ori, which has much less far-infrared excess, shows little
sign of being extended at K in interferometric measurements, and matches
the predictions of accretion disk models quite well \citep{malbet05}. 

In estimating the amount of
solid angle that the envelope must span from the observed long-wavelength
emission, we have included only that emission in the IRS range, i.e.
for wavelengths $\leq 35$ $\mu$m.  Given that the long-wavelength SEDs of
V1057 Cyg and V1515 Cyg are nearly flat, this is clearly an underestimate
of the true total excess emission.  (We do not use the IRAS fluxes because
they very likely include extended, non-envelope emission due to the large
beam sizes.)  If the SEDs were to remain flat out to $\sim 100$ $\mu$m, as in
many Class I sources, the total envelope excesses and thus the inferred solid angle
subtended by the envelope could be significantly larger.  Longer-wavelength
observations, perhaps with MIPS or SOFIA, could help considerably in the characterization of
the geometry of the circumstellar dust in these objects.

\subsection{Envelope model}

\begin{figure}
\plotone{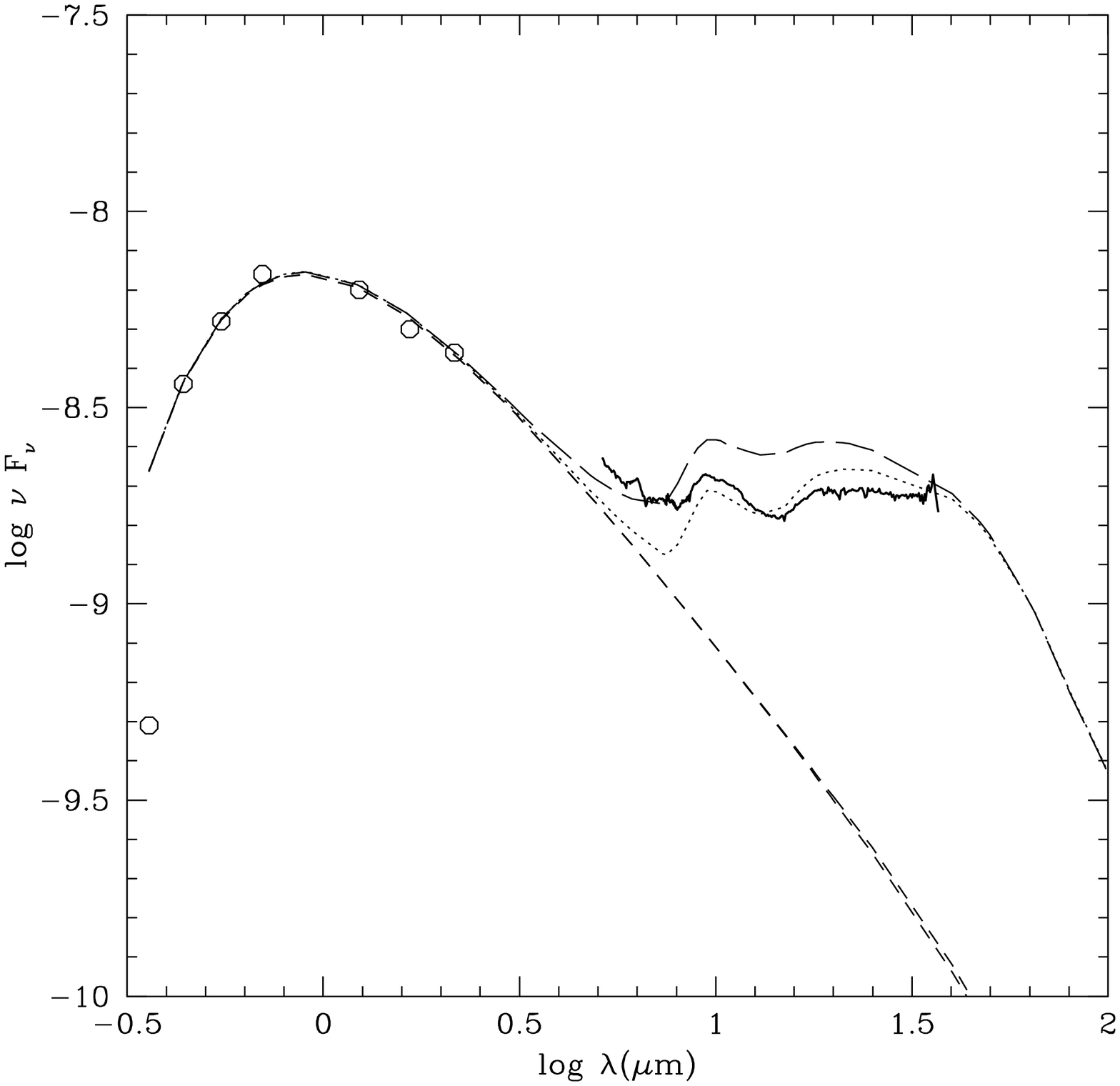}
\caption{Envelope plus steady accretion disk models of V1057 Cyg from \citet{hartmann94} and
\citet{hartmann96b}. The contribution of the disk alone (a series of annuli of blackbodies
at a range of temperatures) is represented by the short dashed curve.  
The dotted curve
shows the fluxes for a model with $\eta = 3.0$, $\rho_1 =
10^{-13} {\rm g cm^{-3}}$, $R_c = 20$ AU, $r_2 = 7000$ AU, $r_1 = 5$ AU, and
$z_2 = 4000$ AU.   
This model fits the long-wavelength
IRS range reasonably well but has too little flux shortward of 10 $\mu$m.
The long-dashed curve is for a model in which the envelope is brought closer in
($r_1 = 3$ AU and $R_c = 10$ AU; the other model parameters are identical to the
previous case).  This model is too bright in most of
the IRS range, although it matches the short-wavelength end better.}
\label{fig7}
\end{figure}

\begin{figure}
\epsscale{0.75}
\plotone{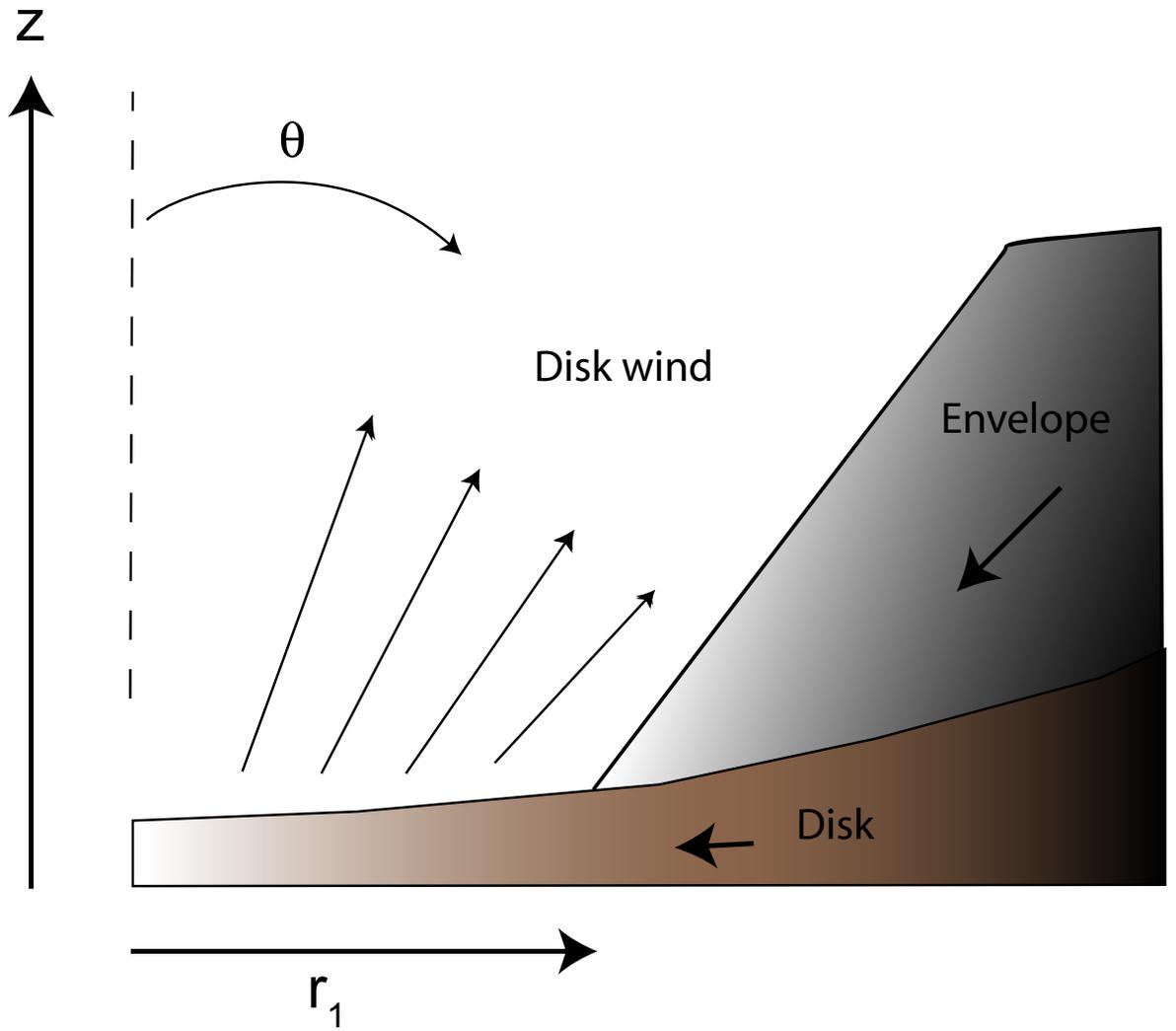}
\caption{A schematic of the steady accretion disk plus envelope model from Figure \ref{fig7}.}
\label{fig8}
\end{figure}

To explore the possible parameters of remnant envelopes, we explore
an approximate calculation and compare it with the observations of V1057 Cyg. 
The models shown in Figure \ref{fig7} combine a simple steady accretion
disk assuming blackbody radiation, plus a flattened envelope with a polar
hole in the density distribution, similar to that used in \citet{osorio03}.  The flattened density
of the envelope is consistent with gravitational collapse of a
sheet beginning in hydrostatic equilibrium as discussed in \citet{hartmann94}
and \citet{hartmann96b}, referred to as $\eta$ models.  The
model parameters include $\eta = R_{outer}/H$, the ratio
of the envelope's outer radius to the scale height \citep{hartmann96b};
$\rho_1$, the density of the infalling envelope at 1 AU given a 
spherical non-rotating envelope; $R_c$, the centrifugal radius, 
the innermost radius on the plane of the disk on which infalling material 
can be deposited.  In addition, we include a polar hole cut out of the
density distribution whose boundary is given by 
\begin{equation}
z = a (r_2 - r_1)\,,
\end{equation}
where $z$ is the height above the equatorial plane and $r_2$ and $r_1$ are
radial distances; $r_1$ signifies the position where the envelope edge
intersects the disk.  The system is then viewed at inclination $i$.
A schematic of this model is displayed in Figure \ref{fig8}.

In order to calculate the temperature structure of the envelope, we
assume radiative equilibrium, using a single central source.  We
calculate the temperature of each spherical shell using a density
averaged over all angles \citep{kenyon93}.  The
luminosity of the central source is given by the steady accretion
models in table \ref{tbl6}.  The dust destruction radius is taken to be at
$\sim$ 1400 K, corresponding to the sublimation temperature
of silicates in a low-density environment \citep{dalessio96}.  The
specific intensity at each wavelength is calculated by solving the
radiative transfer equation along rays that thread the envelope, using
the angle-dependent density distribution, as in \citet{kenyon93} and
\citet{hartmann96b}.  The source function is determined with a mean 
intensity from the spherical case. 

Use of the spherical average radiative equilibrium to determine the
temperature distribution is no longer self-consistent when a (bipolar) hole
is present.  We adjust the luminosity of the central source in the
spherical calculation so that the total luminosity of the envelope
integrated over all solid angle is approximately that expected for
the particular envelope geometry heated by a flat disk, as discussed
in the previous section.

The dust in the envelope uses the standard grain size distribution
of the ISM: $n(a)$ $\alpha$ $a^{-3.5}$.  The grain radii are between $a_{min}
 = 0.005$ $\mu$m and $a_{max} = 0.3$ $\mu$m.  We use a mixture of dust grains from
\citet{pollack94}, as well
as a mixture from \citet{draine84} known as ``astronomical silicates.''
The grains are assumed to be spherical and we calculate their absorption
efficiency factor $Q_{abs}$ using Mie scattering code \citep{wiscombe79}.  The
optical properties of the compounds are taken from \citet{warren84}, \citet{begemann94},
and \citet{pollack94}.

In Figure \ref{fig7} we compare the results of two model envelope calculations
to the observations of V1057 Cyg, for $i= 45^{\circ}$.  The short dashed curve indicates the
contribution of the steady blackbody accretion disk.  This model does a reasonably good job
for the long-wavelength
IRS range but has too little flux shortward of 10 $\mu$m.
The long-dashed  curve is for a model with the same parameters
except for $r_1 = 3$ AU and $R_c = 10$ AU; this model is too bright in most of
the IRS range though it matches the short-wavelength end better.  We suspect
that if we had included the heating of the disk by the envelope
\citep[e.g.,][]{dalessio97}, the 5 - 8 $\mu$m region of the first model would be in
much better agreement with the observations.

Given the complexity of the situation, the envelope parameters derived
for the envelope are not definitive but merely suggestive that envelope models
can account for the long-wavelength IRS fluxes in V1057 Cyg and V1515 Cyg,
as suggested by KH91.  Further progress on this question will require
full two-dimensional axisymmetric radiative transfer calculations.

\subsection{Envelopes and outbursts}

The envelopes we have suggested for V1057 Cyg and V1515 Cyg 
apparently cover modest solid angles as seen from the central sources.
This implies that the (bipolar) outflow holes driven by the winds of
these objects have a wide opening angle.  Theories of 
magnetocentrifugally-driven winds from
young stellar objects suggest that such winds might have a significant
opening angle but have a strong dependence of ram pressure on distance
from the rotation axis, such that the jets frequently seen in such
objects constitute not the entire flow but merely the densest and
most energetic part of a broader flow \citep[e.g.,][]{shu95,matzner99}.

During much of the protostellar
phase, the accretion rates are small; since the mass loss rate tends to track
the accretion rate \citep[e.g.,][]{calvet98}, the outflow is relatively weak
in this phase.  An FU Ori outburst occurs when 
the mass accretion rate increases, perhaps by as much as three to four
orders of magnitude; in consequence, lower-density, wider-angle portions
of the wind can now drive out the infalling envelope.  In this picture,
it is no surprise that the FU Ori objects have wide-angle envelope holes.
V1057 Cyg and V1515 Cyg may be relatively younger, or have had
fewer outbursts, than FU Ori and BBW 76, and thus retain envelope material
at smaller radii/larger covering solid angles.

Of course, each FU Ori outburst is limited in time so that only a small fraction
of a protostellar envelope would be driven off by any given event.  The time
evolution of the envelope geometry in a situation with repeated outbursts
could be quite complicated.  In particular,
some protostellar sources clearly have narrow outflow holes, which may indicate
they have not experienced outbursts, or have particularly large envelope
infall rates and thus higher infall ram pressures. 
The implications of this picture require further exploration. 

If infall and outflow were spherical, ram pressure balance would occur
for the condition

\begin{equation}
\dot{M}_{wind} v_{wind} = \dot{M}_{infall} v_{infall}.
\end{equation}

Consider, for example, V1057 Cyg.  KH91 assumed 
an envelope infall rate of $\sim$ 4 $\times 10^{-6}$ $\msunyr$ in their
envelope model for V1057 Cyg, with an inner radius of about 7 AU.  At that
radius, the infall velocity is (for a 0.5 $M_\sun$ central object) $\sim$ 8 km $s^{-1}$.
With a typical wind velocity of order 300 $\kms$ or more
\citep{croswell87}, a mass loss rate of $\mdot \gtrsim 10^{-7}$ $\msunyr$ would
suffice to blow out the envelope.  As the mass loss rates during peak accretion
rates should be closer to $10^{-5} - 10^{-6} \msunyr$, it is clear that even
a relatively small off-axis wind component will suffice to drive out envelope
material.

\section{Conclusions}

We have presented IRS spectra of five of the best-known FU Ori objects.  One of these sources,
V346 Nor, shows ice absorption features and a large mid-IR excess, indicating that it is
embedded within a cold icy envelope consistent with its very infrared colors.
The other four program objects are moderately-reddened
and show many features in common.  In these objects, the IRS spectra provide
direct evidence of a gaseous accretion disk through the presence of absorption features,
while silicate emission features indicate the presence of outer, cooler dust.

A dusty flared disk probably can explain the long-wavelength dust
emission of two of the objects (FU Ori and BBW 76); this suggests that
FU Ori events can ignite even as the protostellar
envelope is dispersing.  Flared disk
models have difficulties with accounting for the large excesses longward
of 10 microns in V1057 Cyg and V1515 Cyg; instead, we conclude that
an outer dusty envelope is necessary to explain the observations, as
originally suggested by KH91, and which can provide the extended structure in
scattered light detected in near-infrared interferometry by \citet{millan06}.
One important feature of the envelope model
is that it requires an outflow hole with a large opening angle; we suggest that
such a wide outflow-driven cavity is a result of the high mass loss rate
accompanying rapid accretion in the FU Ori outburst state.  Further
near-infrared interferometry with better imaging properties could
confirm our picture of the envelopes around V1057 Cyg and V1515 Cyg.  This
would help support the overall picture of FU Ori outbursts occurring
during the late phases of protostellar collapse.

\acknowledgements

The authors thank Alice Quillen and Judy Pipher for many helpful conversations during the data reduction and
analysis. This work is based on observations made with the {\it Spitzer Space Telescope}, which
is operated by the Jet Propulsion Laboratory, under NASA contract 1407.  Support for
this work was provided by NASA through contract 1257184 issued by JPL/Caltech and
through the {\it Spitzer} Fellowship Program, under award 011 808-001.
This research was supported in part by
Jet Propulsion Laboratory (JPL) contract 960803 to Cornell University,
and Cornell subcontracts 31419-5714 to the University of Rochester.  
The research of LH and NC is supported in part by NASA grant NAG5-13210.
In this work we make use 
of data products from the Two Micron All Sky Survey, which is a joint project of the 
University of Massachusetts and the Infrared Processing and Analysis Center/California 
Institute of Technology, funded by the National Aeronautics and Space Administration 
and the National Science Foundation.  This research has also made use of the SIMBAD database,
operated at CDS, Strasbourg, France.

{}

\clearpage

\begin{deluxetable}{l r r r}
\tabletypesize{\scriptsize}
\tablecaption{Observations Times and AOR IDs\label{tbl1}}
\tablewidth{0pt}
\tablehead{
\colhead{Object} & \colhead{AORID}  & \colhead{Date}
& \colhead{JD}}
\startdata
V1057 Cyg & 3570176 & 2003-12-15 & 2452988 \\
V346 Nor  & 3570688 & 2004-02-27 & 2453063 \\
FU Ori    & 3569920 & 2004-03-04 & 2453068 \\
BBW 76    & 3571200 & 2004-04-14 & 2453110 \\
V1515 Cyg & 3570432 & 2004-05-11 & 2453137 \\
\enddata

\tablecomments{Spitzer-IRS observation log of the FU Ori objects.}
\end{deluxetable}

\begin{deluxetable}{l r r r r}
\tabletypesize{\scriptsize}
\tablecaption{V1057 Cyg in 2004\label{tbl2}}
\tablewidth{0pt}
\tablehead{
\colhead{JD} & \colhead{V}  & \colhead{U-B}
& \colhead{B-V} & \colhead{V-R}}
\startdata
53185.2727 & 12.361 &       & 1.875 & 1.737 \\
53186.2465 & 12.368 &       & 1.890 & 1.755 \\
53187.2281 & 12.364 &       & 1.853 & 1.749 \\
53189.2199 & 12.374 &       & 1.932 & 1.764 \\
53190.2565 & 12.388 &       & 1.980 & 1.772 \\
53195.2785 & 12.444 & 1.484 & 1.892 & 1.769 \\
53196.2637 & 12.469 & 0.986 & 1.875 & 1.770 \\
53205.4167 & 12.463 &       & 1.906 & 1.798 \\
53227.4443 & 12.402 & 1.804 & 1.958 & 1.759 \\
53228.3340 & 12.451 & 1.861 & 1.918 & 1.790 \\
53229.3367 & 12.450 & 2.266 & 1.942 & 1.789 \\
53230.2871 & 12.429 & 0.706 & 1.900 & 1.783 \\
53231.3084 & 12.441 & 1.858 & 1.873 & 1.795 \\
53232.2428 & 12.408 &       & 1.939 & 1.759 \\
53233.2331 & 12.412 & 1.106 & 1.904 & 1.788 \\
53234.2302 & 12.366 & 1.665 & 1.976 & 1.767 \\
53235.3605 & 12.426 &       & 1.939 & 1.736 \\
53236.3604 & 12.466 & 1.190 & 1.906 & 1.776 \\
53237.3542 & 12.495 & 1.418 & 1.876 & 1.806 \\
53238.2251 & 12.483 & 1.123 & 1.909 & 1.788 \\
53240.2332 & 12.474 & 1.123 & 1.907 & 1.779 \\
53241.3390 & 12.477 & 2.277 & 1.886 & 1.775 \\
53242.3228 & 12.499 & 1.616 & 1.956 & 1.793 \\
53243.3501 & 12.488 &       & 1.935 & 1.773 \\
53250.2696 & 12.503 &       & 1.937 & 1.821 \\
53251.2416 & 12.469 & 1.555 & 1.943 & 1.792 \\
53252.2603 & 12.480 &       & 1.891 & 1.789 \\
53253.2421 & 12.458 &       & 1.923 & 1.771 \\
53254.2550 & 12.487 & 1.032 & 1.933 & 1.794 \\
53255.2796 & 12.495 & 2.030 & 1.887 & 1.771 \\
53256.2439 & 12.491 &       & 1.885 & 1.794 \\
53258.2582 & 12.480 & 1.590 & 1.928 & 1.767 \\
53259.3363 & 12.460 &       & 1.912 & 1.774 \\
53260.2522 & 12.513 & 1.340 & 1.882 & 1.782 \\
53261.2235 & 12.485 & 1.958 & 1.905 & 1.767 \\
53262.2218 & 12.478 & 2.552 & 1.906 & 1.759 \\
53263.2217 & 12.446 & 1.384 & 1.952 & 1.756 \\
53263.2386 & 12.388 &       & 2.096 & 1.747 \\
53264.2634 & 12.487 & 2.958 & 1.914 & 1.773 \\
53264.2891 & 12.527 &       & 2.070 & 1.768 \\
53266.1891 & 12.513 & 1.854 & 2.016 & 1.766 \\
53266.2446 & 12.511 & 1.151 & 1.894 & 1.784 \\
53267.2124 & 12.482 & 1.375 & 1.909 & 1.771 \\
53267.2236 & 12.488 & 1.088 & 2.032 & 1.742 \\
53268.1896 & 12.537 & 1.237 & 1.975 & 1.785 \\
53270.2043 & 12.523 &       & 1.904 & 1.763 \\
53271.2405 & 12.537 &       & 1.921 & 1.792 \\
53272.2170 & 12.486 &       & 1.939 & 1.787 \\
53273.2196 & 12.490 &       & 1.951 & 1.765 \\
53289.1960 & 12.600 & 2.405 & 2.013 & 1.802 \\
53291.2346 & 12.644 &       & 1.955 & 1.835 \\
53292.2314 & 12.600 & 1.860 & 1.965 & 1.833 \\
53310.1500 & 12.771 &       & 2.041 & 1.884 \\
\enddata

\tablecomments{The UBVR photometry for V1515 Cygni, FU Orionis and V1057 Cygni,
was observed by Mansur Ibrahimov at Maidanak Observatory in 2004; the average magnitude
is shown here. The UBVR photometry for BBW 76 is from \citet{reipurth02} and has
been projected to 2004, using linear decline rates from the data (see text): 
U - 0.003 mag $yr^{-1}$, B - 0.014 mag $yr^{-1}$, V - 0.023 mag $yr^{-1}$, R - 0.026 mag $yr^{-1}$.
Also included are estimated visual extinction correction factors for each object.}
\end{deluxetable}

\begin{deluxetable}{l r r r r}
\tabletypesize{\scriptsize}
\tablecaption{FU Orionis in 2004\label{tbl3}}
\tablewidth{0pt}
\tablehead{
\colhead{JD} & \colhead{V}  & \colhead{U-B}
& \colhead{B-V} & \colhead{V-R}}
\startdata
53278.4601 & 9.654 & 0.682 & 1.361 & 1.169 \\
53280.4504 & 9.629 & 0.735 & 1.342 & 1.159 \\
53281.4718 & 9.639 & 0.799 & 1.349 & 1.173 \\
53283.4811 & 9.682 & 0.757 & 1.349 & 1.194 \\
53284.4826 & 9.677 & 0.748 & 1.347 & 1.173 \\
53299.4914 & 9.676 & 0.787 & 1.346 & 1.183 \\
53309.4760 & 9.659 &       & 1.350 & 1.176 \\
53310.4598 & 9.661 &       & 1.338 & 1.175 \\
53311.4993 & 9.652 &       & 1.373 & 1.175 \\
\enddata

\tablecomments{Ground based data for FU Ori.}
\end{deluxetable}

\begin{deluxetable}{l r r r r}
\tabletypesize{\scriptsize}
\tablecaption{V1515 Cyg in 2004\label{tbl4}}
\tablewidth{0pt}
\tablehead{
\colhead{JD} & \colhead{V}  & \colhead{U-B}
& \colhead{B-V} & \colhead{V-R}}
\startdata
53184.2758 & 12.973 & 0.620 & 1.599 & 1.509 \\
53185.2461 & 12.990 &       & 1.597 & 1.531 \\
53186.2267 & 12.936 &       & 1.563 & 1.485 \\
53187.2125 & 12.949 &       & 1.566 & 1.477 \\
53189.2030 & 12.926 &       & 1.764 & 1.502 \\
53190.2430 & 12.957 &       & 1.741 & 1.536 \\
53193.2865 & 12.950 & 1.211 & 1.734 & 1.467 \\
53195.2580 & 12.866 &       &       & 1.483 \\
53196.2507 & 12.856 & 1.045 & 1.642 & 1.481 \\
53205.3837 & 13.048 & 1.056 & 1.634 & 1.523 \\
53224.3872 & 13.018 & 0.722 & 1.650 & 1.542 \\
53226.4111 & 13.062 & 1.302 & 1.640 & 1.530 \\
53227.3657 & 13.046 & 1.560 & 1.660 & 1.524 \\
53228.3002 & 13.052 & 1.228 & 1.688 & 1.518 \\
53229.2663 & 13.073 & 1.011 & 1.674 & 1.506 \\
53251.2222 & 13.054 &       & 1.695 & 1.535 \\
53252.2431 & 13.115 &       & 1.642 & 1.586 \\
53253.2262 & 13.116 &       & 1.670 & 1.561 \\
53254.2282 & 13.118 &       & 1.680 & 1.521 \\
53255.2192 & 13.175 &       & 1.637 & 1.543 \\
53256.2161 & 13.160 &       & 1.649 & 1.508 \\
53258.2446 & 13.199 & 1.185 & 1.665 & 1.562 \\
53259.3272 & 13.137 & 1.705 & 1.680 & 1.528 \\
53260.2360 & 13.121 & 2.027 & 1.697 & 1.543 \\
53261.2048 & 13.054 & 2.366 & 1.733 & 1.545 \\
53262.2066 & 13.066 & 1.968 & 1.645 & 1.545 \\
53263.2063 & 13.049 & 0.660 & 1.689 & 1.549 \\
53264.2492 & 13.064 & 1.191 & 1.683 & 1.520 \\
53266.2246 & 13.071 & 1.865 & 1.636 & 1.564 \\
53267.1977 & 13.033 & 1.379 & 1.660 & 1.505 \\
53268.2515 & 12.977 &       & 1.664 & 1.489 \\
53270.1930 & 12.952 &       & 1.624 & 1.507 \\
53271.2320 & 12.963 &       & 1.643 & 1.490 \\
53272.1869 & 13.011 &       & 1.586 & 1.527 \\
53273.1885 & 13.039 &       & 1.654 & 1.530 \\
53289.1777 & 13.108 & 1.347 & 1.632 & 1.536 \\
53292.1552 & 13.042 & 0.912 & 1.615 & 1.489 \\
\enddata

\tablecomments{Ground-based data for V1515 Cyg.}
\end{deluxetable}

\begin{deluxetable}{l r r r r r}
\tabletypesize{\scriptsize}
\tablecaption{Estimated 2004 Brightness Levels and Visual Extinction\label{tbl5}}
\tablewidth{0pt}
\tablehead{
\colhead{Object} & \colhead{U(mag)}  & \colhead{B(mag)} &  \colhead{V(mag)} & \colhead{R(mag)} & \colhead{$A_V$}}
\startdata
V1057 Cyg & 17.00 & 14.70 & 12.65 & 10.85 & 3.7 \\
FU Ori    & 11.81 & 11.01 & 9.67 & 8.49 & 1.9 \\
BBW 76    & 14.23 & 14.04 & 12.59 & 11.81 & 2.2 \\
V1515 Cyg & 16.50 & 14.74 & 13.11 & 11.57 & 3.2 \\
\enddata

\tablecomments{Current UBVR photometry and uncorrected for reddening, and assumed reddening corrections for the FU Ori objects.  The average values for V1057 Cyg, V1515 Cyg, and FU Ori are computed from tables \ref{tbl2}, \ref{tbl3}, and
\ref{tbl4}.  The values for BBW 76 are extrapolations from older data, as explained in the text.}
\end{deluxetable}

\begin{deluxetable}{l r r r r r}
\tabletypesize{\scriptsize}
\tablecaption{Steady Accretion Model Parameters\label{tbl6}}
\tablewidth{0pt}
\tablehead{
\colhead{Object} & \colhead{d(kpc)}  & \colhead{$L/L_{\odot}$}
& \colhead{$T(max)$} & \colhead{$R_i/R_{\odot}$} 
& \colhead{$M \dot{M}/(10^{-4} M_{\odot}^2 {\rm yr^{-1}})$}} 
\startdata
FU Ori    & 0.5    & 466 & 7710  & 5.0 & 1.5  \\
V1057 Cyg & 0.6    & 170 & 6590  & 3.7 & 0.45  \\
V1515 Cyg & 1.0    & 177 & 7710  & 3.1 & 0.35  \\
BBW 76    & 1.8    & 287 & 7710  & 3.9 & 0.72  \\
\enddata

\tablecomments{Distances are from \citet{hartmann96a}, except for
BBW 76, which is taken from \citet{reipurth02}.
Note: if V1057 Cyg and V1515 Cyg are observed pole-on, then
the true accretion luminosities may be a factor of up to two smaller,
with a reduction of $2^{1/2}$ in the inner radius and a reduction
of $2^{3/2}$ in $M \dot{M}$.}
\end{deluxetable}

\end{document}